\documentclass{ws-procs9x6}
\usepackage{subfigure}
\usepackage{graphics}
\usepackage{epsf,amsmath}
\usepackage{amssymb,epsfig,floatflt}

\begin{document}

%\begin{titlepage}{GLAS-PPE/2005-??} {22$^{\underline{\rm{nd}}$ November 
%2005}
\title{Particle Production and Fragmentation at HERA}

\author{D.~H. SAXON}

\address{Faculty of Physical Sciences, \\ 
University of Glasgow, \\
Glasgow, G12 8QQ, Scotland\\
E-mail: d.saxon@physics.gla.ac.uk}  

\maketitle

%\abstracts{
%This is where the abstract should be placed. It should consist
%of one paragraph giving a concise summary of the material in 
%the article below.  Replace the title, authors, and addresses 
%within the curly brackets with your own title, authors, and
%addresses. You may have as many authors and addresses as you 
%like. It is preferable not to use footnotes in the abstract or 
%the title; the acknowledgments for funding bodies etc. are to 
%be placed in a separate section at the end of the text. Please see 
%the appendices too.}

%\begin{abstract}
\abstracts{
Recent results from HERA are presented on a range of topics: charged 
multiplicities, 
production of non-strange mesons and strange particles, charm 
fragmentation, baryons decaying to strange particles, antideuteron 
production, Bose-Einstein correlations, and new interpretations of 
results 
on prompt photon production in DIS.}

%\vspace{0.5cm}
%\begin{center}
%{\em New Trends in HERA Physics 2005}\\
%{\em Ringberg, Oct 2-7, 2005}
%\end{center}
%\end{abstract}

%\newpage
%\end{titlepage}

%\section{Introduction}
%\subsection{Producing the Hard Copy}\label{subsec:prod}
%The hard copy may be printed using the advice given in the file
%{\em procs-readme9x6$\_$2e.pdf}, which is repeated in this section. 
%Total there are seven files given.\footnote{You can obtain these files 
%from our WWW pages at:

%{\sf http://www.wspc.com.sg/style/proceedings\_style.shtml}}

This article reports on recent experimental results from H1 and ZEUS, 
together with new theoretical ideas in the case of prompt photon
production. The topics covered are

%\begin{enumerate}
\begin{itemize}

\item{Charged Multiplicities in DIS and diffractive DIS (DDIS). The use 
of 
$W$ and $E_{current}$ and a unified approach to DIS and DDIS.}

\item{Inclusive photoproduction of non-strange mesons. Universal rate 
curve as a function of $(p_T + m)$.}

\item{Strange Particle Production and polarisation, looked at as a 
function 
of $p_T, \eta , Q^2, x$.}

\item{Charm fragmentation. Universality in DIS, photoproduction and $e^+ 
e^-$ annihilation.}

\item{Baryons decaying to strange particles.}

\item{Antideuteron production. $dE/dx$. Coalescence models.}

\item{ $KK$ Bose-Einstein correlations. Comparison to LEP, $f_0(980)$ 
issues.}

\item{Prompt photons in DIS. First measurement of 
$\gamma_p(x,Q^2)$?}

\end{itemize}

%\item {\em procs-readme9x6$\_$2e.pdf} --- the preliminary guide.

%\item {\em procs-instruction9x6$\_$2e.pdf} --- general instructions 
%for authors.

%\item{\em procs-fig1.eps} --- the figure/image file.

%\item{\em rotating$\_$pr.sty} --- sty file for landscape figures and 
%tables.

%\item {\em ws-procs9x6.cls} --- the class file that provides the higher
%level latex commands for the proceedings. Don't change these parameters.

%\item {\em ws-procs9x6.tex} --- the main text. 

%\item {\em ws-procs9x6.pdf} --- sample pages of the above text.
%\end{enumerate}

\section{Charged Particle Multiplicities in DIS and DDIS.}
Historically, charged particle multiplicities have been measured in DIS 
in the current region of the Breit frame - defined as the frame in which 
there is no energy transfer in the collision. In zeroth order perturbative 
QCD this corresponds to a quark-parton jet with energy $Q/2$ and can be 
compared to a half-event in $e^+ e^-$ annihilation. Early ZEUS results 
 covered  the range $10 < Q^2 < 4000~ \rm{GeV^2}$ with rather large 
errors at the high end. The results lie mostly on top of $e^+ e^-$ 
annihilation but fall below those at $Q^2 < 50~ \rm{GeV^2}$.
New analyses in the Breit frame using $2E_{current}$ as the scale 
instead of $Q/2$ fix this problem at low energy - see figure 1.

\begin{figure}[ht]
%\epsfxsize=10cm   %width of figure - will enlarge/reduce the figures
%\epsfbox{fig3.eps}
%\figurebox{2cm}{3cm}{} %to have a box alone 
\centerline{\epsfxsize=4.1in\epsfbox{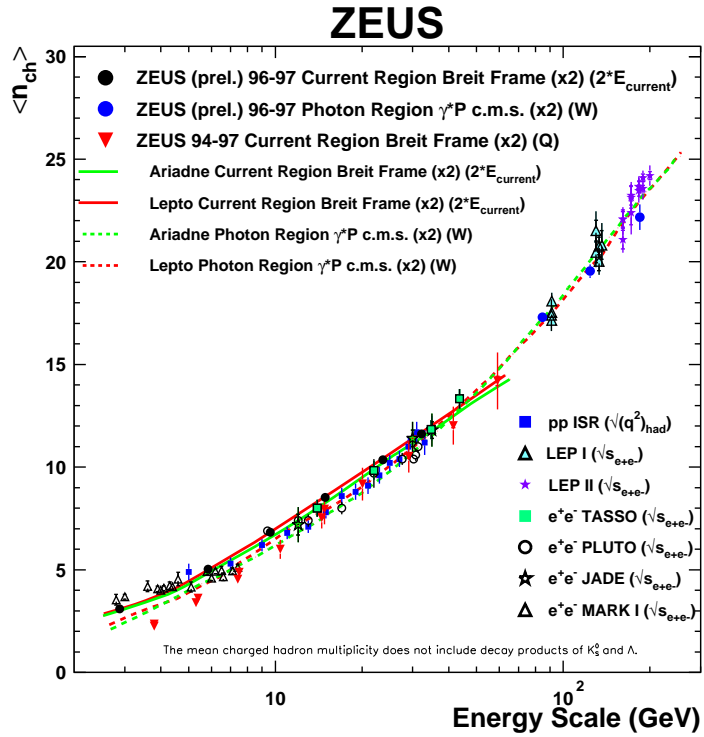}}   
%\centerline{\epsfxsize=4.1in\epsfbox{multiplicity_fig2.eps}}   
\caption{Charged particle multiplicities. \label{mult}}
\end{figure}

Using one hemisphere of the Breit frame offers a rapidity range of only 
$\ln(Q/m)$ where $m$ is the pion mass. Instead we can work in the 
$\gamma^* p$ centre-of-mass frame using the hadronic energy $W$ as the 
scale and access instead a larger rapidity range of $\ln(W/m)$ (Recall 
$W^2 
\simeq Q^2/x$.) These results are also shown in figure 1 and show 
excellent agreement between DIS and $e^+ e^-$, and with calculations 
using {\sc Lepto} and {\sc Ariadne.} 

ZEUS have also looked at an alternative approach, which is to use only the 
best part of the tracker ($20^{\circ} < \theta < 160^{\circ}$) where the
tracking efficiencies are high and well-known and to plot the multiplicity 
(excluding the recoil electron) 
as a function of $M^2_{\rm{eff}} = (\Sigma E)^2 - (\Sigma \mathbf{p})^2$. 
The 
results cover the range $5 < M_{\rm{eff}} < 40~\rm{GeV}$ and agree well 
with
{\sc Ariadne}. There is a weak Bjorken-$x$ dependence across the range 
$0.006 < 
x < 0.1$.

H1 have compared multiplicities in DIS and DDIS. The variable used to 
characterise the DDIS event is the mass of the diffractively produced 
hadronic system, $M_x$, giving a rapidity range of $\ln(M_x / m)$.
Many comparison plots are made of mean multiplicity versus $W$ or $Q^2$ 
for different $M_x$ ranges. As $M_x$ increases, 
the DDIS mean multiplicities, and their rapidity spectra, smoothly 
approach the DIS values from below. This is as one might  expect naively 
in a colour-string approach where what matters is the number of units of 
rapidity of colour-string available to fragment into hadrons.

\section{Non-strange inclusive meson photoproduction}
H1 have analysed the $\pi^+ \pi^-$ mass spectrum in photoproduction. Mass 
peaks for $\rho^0, f_0$ and $f_2$ are identified over a large and steeply 
falling background. They plot (see figure 2) a universal curve of 
spin-weighted rate against $p_T + m$. Within the discrimination of a 
log-log 
plot, these mesons and $\eta$ and $\pi^+$ lie on the same line and 
follow a power law over six orders of magnitude.

\begin{figure}[ht]
%\epsfxsize=10cm   %width of figure - will enlarge/reduce the figures
%\epsfbox{fig3.eps}
%\figurebox{2cm}{3cm}{} %to have a box alone 
%\centerline{\epsfxsize=4.1in\epsfbox{H1prelim-03-037.fig5.epsi}}   
\centerline{\epsfxsize=4.1in\epsfbox{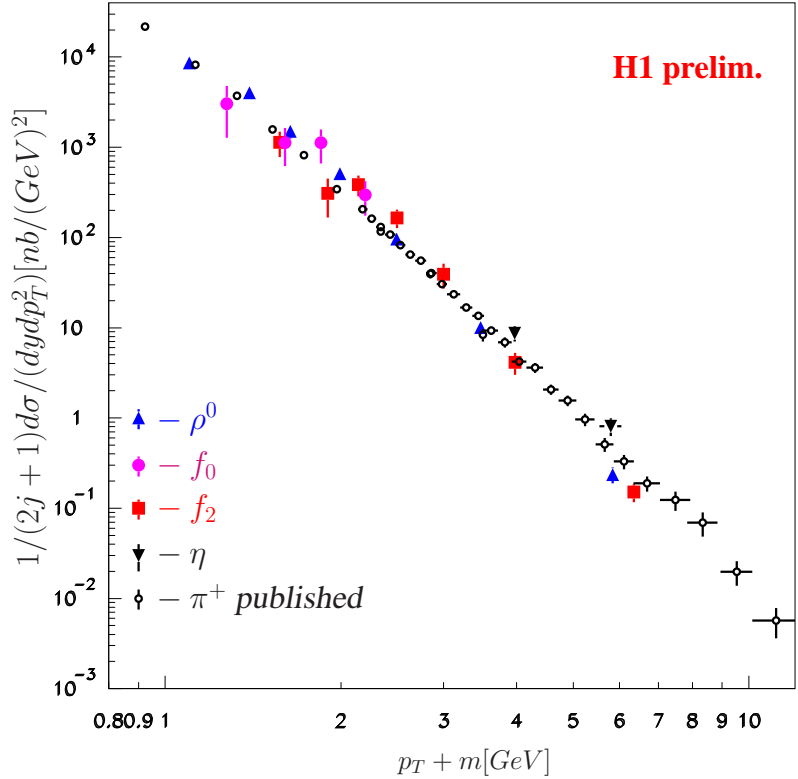}}   
\caption{Meson production rates. \label{mesonrates}}
\end{figure}

\section{$K^0, \Lambda^0$ and $\overline{\Lambda}^0$ production in DIS.}
\begin{figure}[ht]
%\epsfxsize=10cm   %width of figure - will enlarge/reduce the figures
%\epsfbox{fig3.eps}
%\figurebox{2cm}{3cm}{} %to have a box alone 
%\centerline{\epsfxsize=4.1in\epsfbox{lambda_fig6.eps}}   
\centerline{\epsfxsize=4.1in\epsfbox{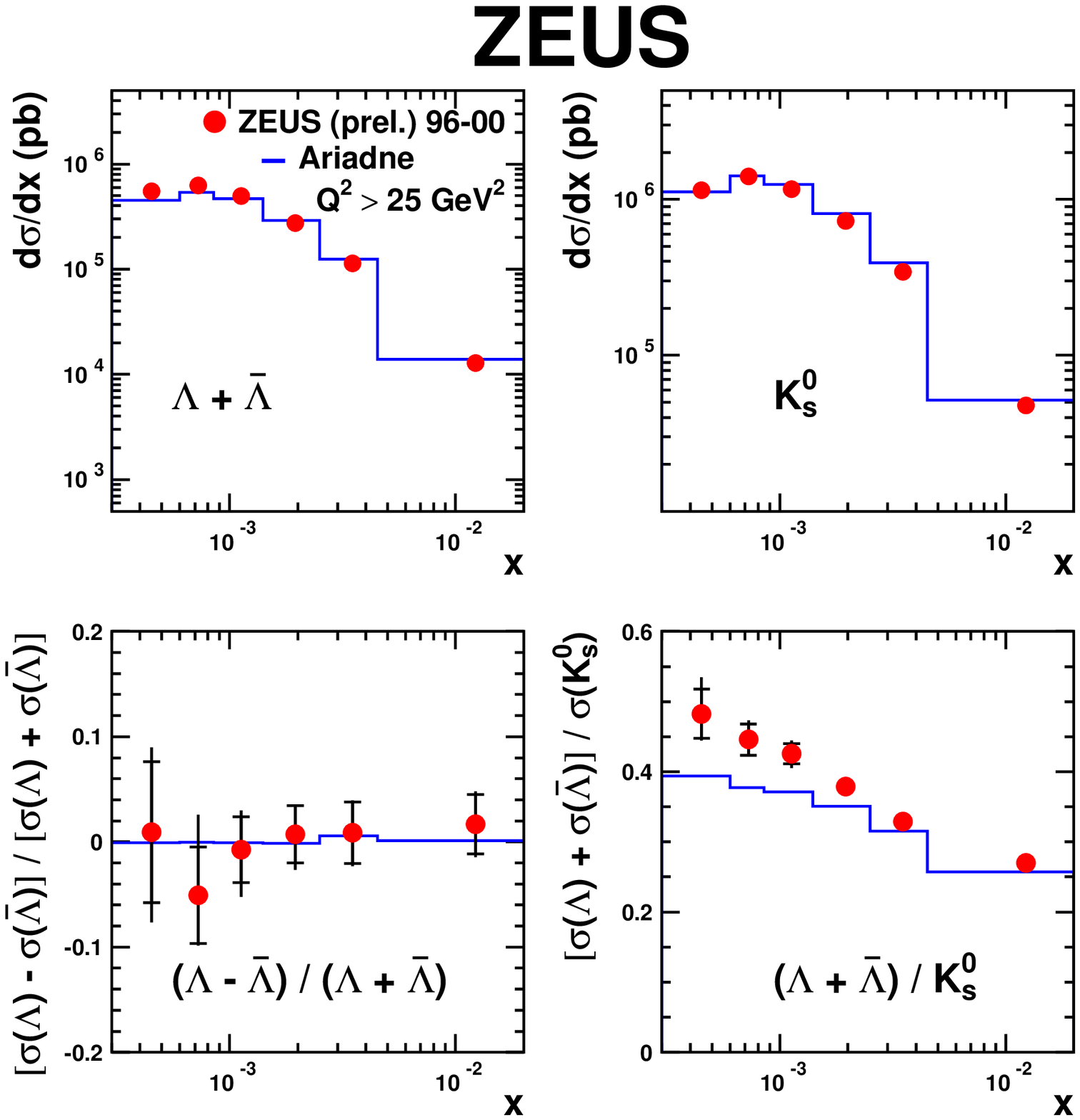}}   
\caption{Strange particle production compared to predictions of 
{\sc Ariadne}. 
\label{lambda}}
\end{figure}

ZEUS present clean mass peaks for these strange particles and plot the 
$p_T, \eta, Q^2$ and $x$ distributions for $K_s^0$,
$\Lambda^0 + \overline{\Lambda}^0$, and the ratios 
$(\Lambda^0 + \overline{\Lambda}^0)/K_s^0$, and
$(\Lambda^0 - \overline{\Lambda}^0)/(\Lambda^0 + \overline{\Lambda}^0)$
for $Q^2 > 25~\rm{GeV^2}$. The error bars, (see figure 3,) are small and 
there
is mostly very pleasing agreement with the predictions of {\sc Ariadne}. 
{\sc Ariadne} 
tends to overestimate the $K^0$ production rate, suggesting that a
lower value of $s/u \simeq 0.22$ may be appropriate. No signficant
difference is found between
$\Lambda^0$ and  $\overline{\Lambda}^0$. The $\Lambda / K$ ratio
 is underestimated by {\sc Ariadne} for $x < 0.003$, as shown in figure 
\ref{lambda}.

$\Lambda^0$ and  $\overline{\Lambda}^0$ polarisations are measured using 
the $\Lambda \rightarrow \pi^- p$ decay as analyser. In the $\Lambda^0$ 
rest frame the decay angular distribution with respect to a chosen axis is 
proportional to $(1 + \alpha P \rm{cos} \theta )$ with $\alpha = 0.642$.
Using  unpolarised $e^\pm$ beams no significant polarisations are observed 
either normal to the production plane (with $6\%$ errors) or along the 
$\Lambda$ flight path (with $12\%$ errors). They have thus demonstrated 
the ability to measure these polarisations, with possible application to 
polarised $e^\pm$ running at HERA-II.

\section{Charm fragmentation.}
Results from H1 and ZEUS measure the production rates of
$D^{*+}, D^+, D^0, D^+_s$ and $\Lambda^+_c$. $D^{*+}$ are measured by the
traditional decay chain $D^{*+} \rightarrow D^0 \pi^+,~D^0 \rightarrow K^- 
\pi^+$. H1 use vertex detector signatures to isolate the decays $D^+ 
\rightarrow K^- \pi^+ \pi^+$ and $D^0 \rightarrow K^- \pi^+$ with 
relatively small backgrounds. They also use vertex tagging to  measure 
$D^\pm_s$ production via the decay 
chain $D^\pm_s \rightarrow \phi \pi^\pm,~\phi \rightarrow K^+ K^-$. 
ZEUS observe the same channel and also
find a signal of $(1440 \pm 220)$ $\Lambda^+_c$ events on a background of 
some 20000 events in the $p K^- \pi^+$ decay channel.

This allows them to compare fragmentation fractions such as $f(c 
\rightarrow D^0 )$ in photoproduction, DIS (using 
both H1 and ZEUS data) and $e^+ e^-$ annihilation. There is good agreement
on the fractions in all cases, though there is a hint that ZEUS find a 
somewhat lower value for $f(c \rightarrow D^{*+})$ than $e^+ e^-$ 
annihilation. 

\section{Baryons decaying to strange particles.}
The search for pentaquarks has revived interest in baryons decaying to 
strange particles. We do not cover pentaquark issues here but look at 
other 
particles whose production has now been observed, often small but 
significant signals on large backgrounds. Both ZEUS and H1 identify 
protons 
and antiprotons by $dE/dx$ information, restricted to momenta below 
$1.5~\rm{GeV/c}$. ZEUS demand that charged tracks emerge from the 
primary 
production vertex and identify $K^0_s$ via a $\pi^+ \pi^-$ secondary 
vertex, selecting $p_T (K^0_s) > 0.3~\rm{GeV/c}$ and 
$-1.5<\eta(K^0_s)<1.5$ to use the best part of the tracker. Additional 
cuts exclude Dalitz pairs, $\gamma$-conversions and $\Lambda^0$ 
candidates. Their $(K^0_s p)$ mass resolution is $2.4~\rm{MeV}$.

\begin{figure}[ht]
%\epsfxsize=10cm   %width of figure - will enlarge/reduce the figures
%\epsfbox{fig3.eps}
%\figurebox{2cm}{3cm}{} %to have a box alone 
\centerline{\epsfxsize=4.1in\epsfbox{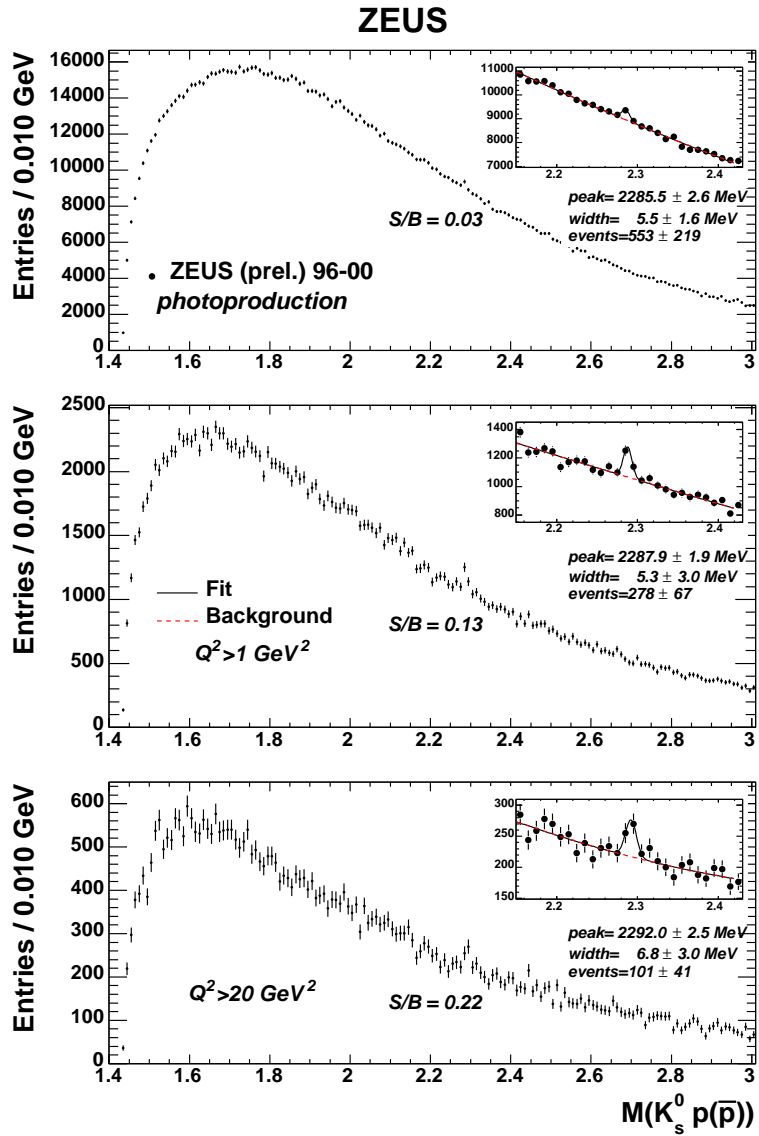}}   
%\centerline{\epsfxsize=4.1in\epsfbox{pentaquarks_fig1.eps}}   
\caption{Observation of $\Lambda_c(2286)$ production. \label{lc}}
\end{figure}

Three data samples are used: photoproduction, DIS for $Q^2 > 
1~\rm{GeV}^2$, and DIS for $Q^2 > 25~\rm{GeV}^2$. The particle 
multiplicities, and hence the backgrounds are highest in photoproduction 
and lowest in the high-$Q^2$ sample. In the $K^0_s p(\overline{p})$ mass 
spectrum, as well as a pentaquark candidate, $\Theta(1530)$, in DIS a 
clear $\Lambda_c (2286)$ is seen in all three samples with a width of 
typically $5.3 \pm 3.0~\rm{MeV}$ (see figure \ref{lc}.) The most prominent 
signal is $(278 \pm 
67)$ events in  low-$Q^2$ DIS. It is seen equally in $K^0_s p$ and
$K^0_s \overline{p}$ ($162 \pm 36$ and $116 \pm 38$ events) and equally at 
forward and backward lab rapidity ($131 \pm 40$ and $145 \pm 34$ events.)
This is consistent with expectations for $\gamma g \rightarrow c 
\overline{c}$ production.

A clear $\Lambda (1520)$ signal is seen equally in both $K^- p$ and $K^+ 
\overline{p}$ mass spectra in all three data samples: in photoproduction 
they observe $(13526 \pm 561)$ events giving $S/B = 0.05$ with errors of 
$0.4~\rm{MeV}$ on both mass and width. Again the production is the same at 
forward and backward rapidity, again consistent with photon-gluon fusion.

In the $\Lambda \pi^\pm$ spectrum one searches for $\Xi,\Sigma^*$ and 
pentaquark production. $\Xi(1320)$ and $\Sigma^* (1385)$ production are 
seen but there is no peak in the $\Theta^+ (1530)$ region. There is a 
$4.6~\sigma$ peak near 1600 MeV in the $Q^2 > 1 \rm{GeV}^2$ sample, which 
might be $\Sigma(1580)$ or $\Sigma(1620)$ production.

Very clean peaks of $(1561 \pm 46)$ events in $\Xi^- \rightarrow \Lambda 
\pi^-$ and the same number in $\overline{\Xi}^+$ decay are seen after cuts 
are made to ensure a clean $\Lambda$ decay vertex. These in turn lead to a 
clear narrow $\Xi^0 (1530)$ signal in the $\Xi^- \pi^+$ (and antiparticle) 
mass spectrum.  

\section{Anti-deuteron production and heavy particle search.}
H1 have used their excellent $dE/dx$ resolution to measure the 
photoproduction 
of $K,p,d,t$ and antiparticles for $\langle W(\gamma p) \rangle = 
200~\rm{GeV}$, $0.2 < p_T/M < 0.7$, $ -0.4<y_{lab}<0.4$, using 
$5.5~\rm{pb}^{-1}$ of data. Clear peaks are 
observed for $K^+,p,d,t,K^-$ and $\overline{p}$ with $7\%$ mass 
resolution, and a clean cluster of 45 antideuterons is observed,
corresponding to a cross section of $(2.7 \pm 0.5 \pm 0.2~\rm{nb})$. There 
are no candidates for heavier negative particles. 

Antideuteron production is beyond standard fragmentation models.
H1 compare their results to $pp$ collisions at the ISR and to $AuAu$ 
collisions in STAR at RHIC as a function of $p_T/M$. Plotting the 
cross-section/total cross-section and the $\overline{d} / \overline{p}$ 
ratio the photoproduction and $pp$ data are in good agreement but the 
$AuAu$ data are much higher. The data are also tested against a
coalescence model, which favours $d$ production if $p$ and $n$ are 
produced very close together. Heavy ion collisions have a much larger 
production volume and hence a much smaller chance of overlap. Again we 
find agreement between $\gamma p$, $pp$ and $pA$ over a wide c.m. energy 
range, but very heavy ions $NeAu$, $AuPt$ and $PbPb$ give much lower 
coalescence probabilities, falling rapidly as the c.m. energy increases.

\section{Bose-Einstein correlations in $K^0_s K^0_s$ and $K^{\pm} 
K^{\pm}$.}

ZEUS have performed a  Bose-Einstein analysis for charged and 
neutral kaons. The correlation function used for two particles is
$R(p_1,p_2) = \rho(p_1,p_2)/\rho(p_1)\rho(p_2)$ . We set $Q_{12} = p_1 - 
p_2$ and fit the data to the form
\begin{center}
$R(Q_{12}) = \alpha (1 + \delta Q_{12})( 1 + \lambda \exp [-r^2 
Q^2_{12}])$
\end{center}
\noindent
where the physics interest lies in the source radius, $r$, and the
incoherence parameter, $(0 < \lambda < 1)$. $R(Q_{12})$ is  measured in 
the data by the double ratio
\begin{center}
$R = [P({\rm data})/P_{mix}({\rm data})]/[ P({\rm MC})/P_{mix}({\rm MC})]$
\end{center}
\noindent
all evaluated at the same $Q_{12}$,
where the suffix $mix$ implies that the two particles are taken from 
different events. The Monte-Carlo has no Bose-Einstein correlations.

Looking at $K^{\pm} K^{\pm}$ in DIS events ($Q^2 > 2~\rm{GeV}^2$) ZEUS 
find $r = (0.57 \pm 0.09 +0.15-0.06)~\rm{fm}$ in good agreement with
LEP and with charged pions. The $\lambda$ value, $(0.31 \pm 0.06 + 
0.09-0.06)$, is less than half that at LEP. The reason is unclear, though 
we note that the fragmentation may be different in the proton region of 
the c.m. frame and that $\phi (1020)$ decay may therefore play a 
different role.

\begin{figure}[ht]
%\epsfxsize=10cm   %width of figure - will enlarge/reduce the figures
%\epsfbox{fig3.eps}
%\figurebox{2cm}{3cm}{} %to have a box alone 
\begin{minipage}{0.44\linewidth}
\begin{center}
\subfigure{
%\centerline{\epsfxsize=4.0in\epsfbox{be_fig3.eps}}   
%\centerline{\epsfxsize=4.0in\epsfbox{be_fig3.eps}}   

%\epsfig{figure = be_fig3.eps,
\epsfig{figure = 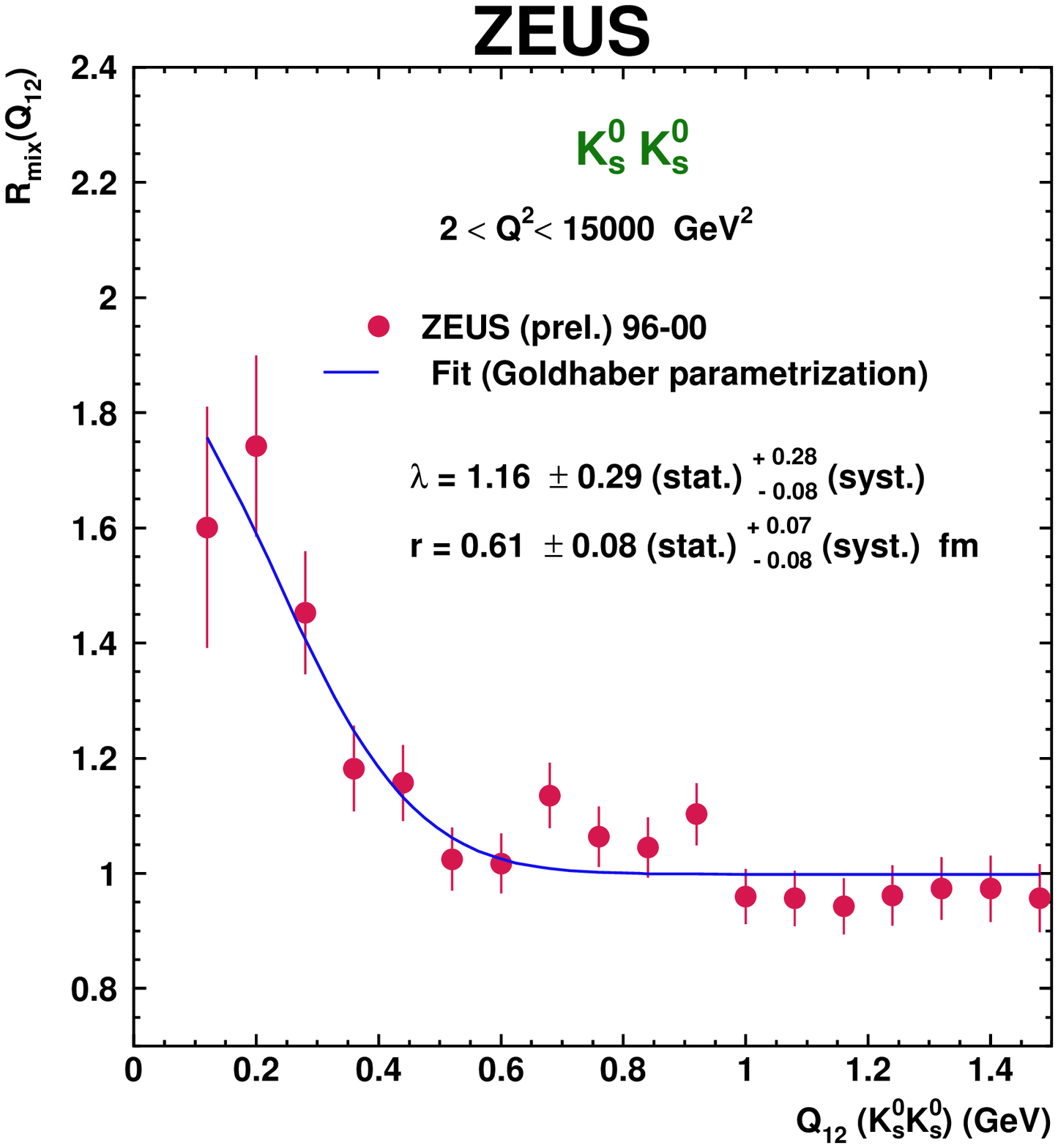,
width={2.2in}, angle=0, clip=}
}
\end{center}
\end{minipage}
\begin{minipage}{0.44\linewidth}
\begin{center}

\subfigure{
%\centerline{\epsfxsize=4.0in\epsfbox{be_fig1.eps}}   
%\epsfig{figure = be_fig1.eps,
\epsfig{figure = 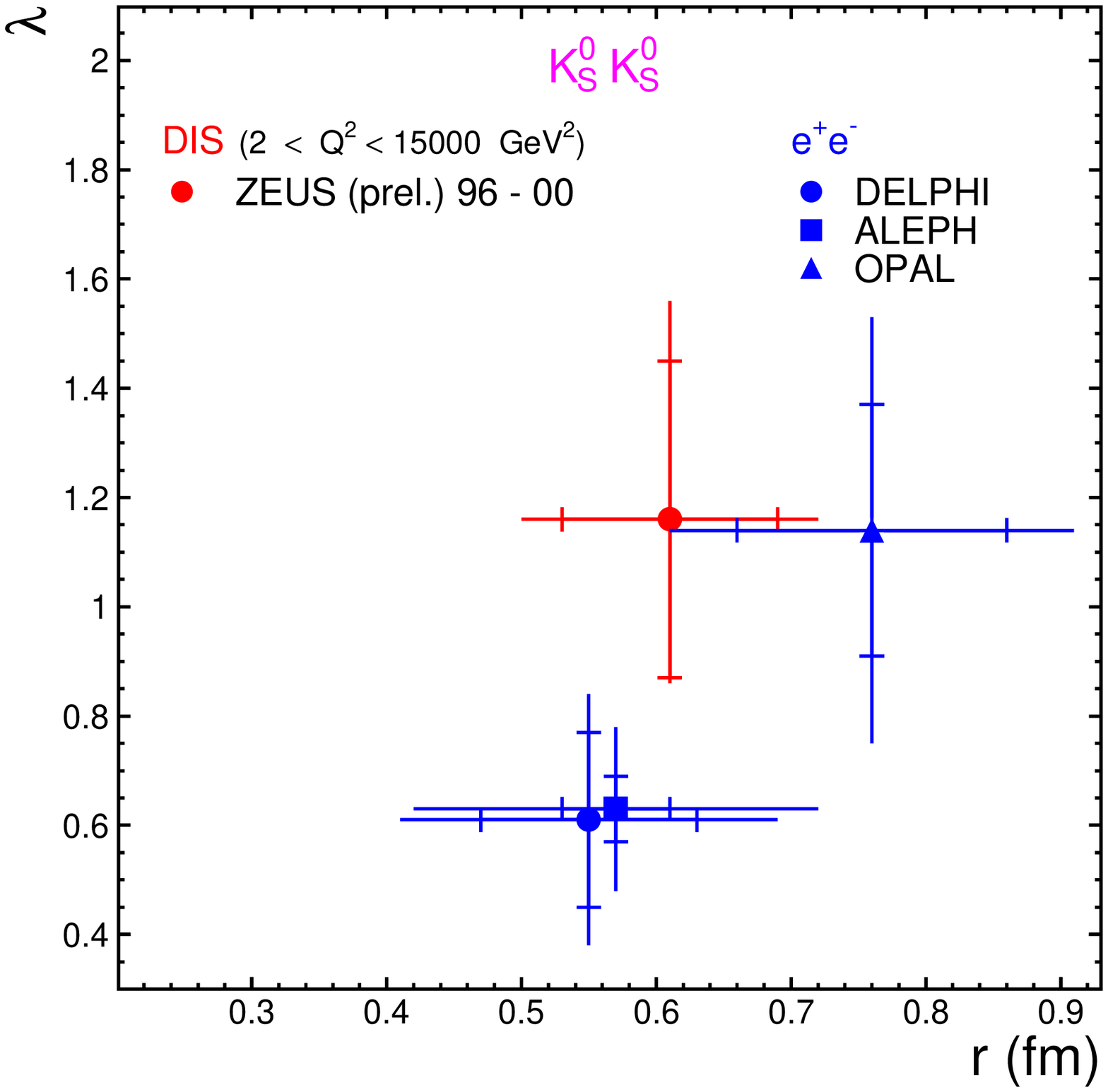,
width={2.2in}, angle=0, clip=}
}
\end{center}
\end{minipage}
\caption{Bose-Einstein correlations. \label{mesons}}
\end{figure}

In $K^0_s K^0_s$ ZEUS find a very similar value for $r$, in contrast with 
LEP results where there is a hierarchy, $r(\pi^\pm) > r(K^\pm) > 
r(\Lambda)$. (See figure 5.) ZEUS find $\lambda \simeq 1.2$, similar to 
OPAL but
well above values from ALEPH and DELPHI who have removed $f_0(980) 
\rightarrow K^0_s K^0_s$ effects. Correction for this may bring the 
results into agreement. 

\section{Prompt photon emission.}
We are looking here for hard photons emitted from the quark line in 
photoproduction or DIS. The experimental definition of photoproduction 
demands that the electron be lost down the beam pipe, so that a photon 
observed in the detector with $p_T > 5~\rm{GeV}$ must be far from it.
We demand that the photon be isolated from all 
other outgoing particles by drawing a cone of radius 1 in $(\Delta \eta, 
\Delta \phi)$ around it and demanding that the electromagnetic cluster 
identified with the photon contain at least $90\%$ of the energy found in 
the isolation cone. In DIS the photon is therefore far from both electron 
and quark.

$\pi^0$ and $\eta^0$ decay to photons provide major backgrounds to prompt 
photon signals. Both ZEUS and H1 have used transverse shower shape to 
discriminate these. In the ZEUS barrel calorimeter ($ -0.7 < \eta < 0.9$) 
the fine segmentation means that photons predominantly illuminate only 1 
$z$-strip, $\pi^0$ decay mostly two strips amd $\eta$ decay fills one to 
many 
strips. A photon signal is extracted statistically from fitting the shape 
distributions.

Photoproduction of prompt photons is rather well described by both 
{\sc PYTHIA} 
(shape OK, normalisation a bit low) and by NLO QCD. H1 measurements of 
$E^\gamma_T, \eta^\gamma$ and $\eta^{\rm{jet}}$ for accompanying jets are 
all described well. Corrections to NLO for multiple interactions (as 
estimated by {\sc PYTHIA}) improve the fit, but there are no surprises.

Deep inelastic scattering turns out to be more of a challenge. ZEUS have 
published measurements of inclusive prompt photon emission and of the 
(photon+jet) final state, and they compared these to two Monte Carlo 
calculations ({\sc PYTHIA~v6.206} and {\sc HERWIG~v6.1}) and to NLO (that 
is 
$O(\alpha^3 \alpha_s)$ calculations by Kramer and Spiesberger. The data
are shown in figure 6. {\sc PYTHIA} and {\sc HERWIG} get the normalisation 
wrong, by factors (in the inclusive case) of 2.4 and 8.3 respectively, and 
{\sc HERWIG} gets too low a mean $Q^2$, while {\sc PYTHIA} gets the slope 
of the 
rapidity spectrum wrong. A similar story holds in the case of prompt 
photon plus one jet.

%figure here

\begin{figure}[ht]
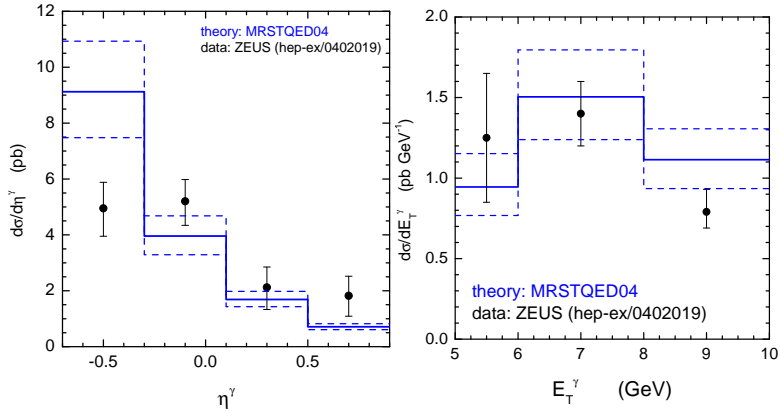

%\epsfxsize=10cm   %width of figure - will enlarge/reduce the figures
%\epsfbox{fig3.eps}
%\figurebox{2cm}{3cm}{} %to have a box alone 
\begin{minipage}{0.44\linewidth}
\begin{center}
\subfigure{
%\centerline{\epsfxsize=4.0in\epsfbox{be_fig3.eps}}   
%\centerline{\epsfxsize=4.0in\epsfbox{be_fig3.eps}}   
%\epsfig{figure = ZEUSptg.epsi,
\epsfig{figure = saxonfig6a.epsi,
width={2.0in}, angle=0, clip=}

}
\end{center}
\end{minipage}
\begin{minipage}{0.44\linewidth}
\begin{center}
\subfigure{
%\centerline{\epsfxsize=4.0in\epsfbox{be_fig1.eps}}   
%\epsfig{figure = ZEUSetag.epsi,
\epsfig{figure = saxonfig6b.epsi,
width={2.0in}, angle=0, clip=}
}
\end{center}
\end{minipage}
\caption{Inclusive prompt photon production in DIS compared to
MRST predictions (absolute normalisation). \label{prph}}
\end{figure}

Comparisons with NLO calculations, only available for the photon plus jet 
case, are more encouraging. The normalisation and mean $Q^2$ are good, 
(within 1.7 S.D. for cross-section) as 
are the photon rapidity and $E_T^{\rm{jet}}$ distribution. But the jet 
rapidity is predicted to be more forward peaked than the data and the 
photon $E_T$ spectrum is more steeply falling than the data. More 
statistics will obviously help to clarify whether these issues are
significant or not.

After these results were published, Martin, Roberts, Stirling and Thorne 
produced a different approach, in which the outgoing hard photon is
emitted by the electron, which in turn scatters off a photon which is a
constituent of the proton, produced by QED/QCD evolution of the proton 
structure function. The high $Q^2$ in the event (averaging 
$85~\rm{GeV}^2$) in the data is carried by the exchanged electron.
MRST calculate the photon structure of the proton $x \gamma_p(x,Q^2)$ to 
be about 0.03 at $x \simeq 0.005$, (i.e. about 300 times  lower that the 
gluon content. This in turn gives a cross-section prediction similar to 
the data (well within 1 S.D.) The predictions for inclusive photon 
production, (they do not calculate photon+jet,) are overlaid on the data 
in figure 6.

The upshot is in some sense intriguing. Both the NLO and the structure 
approach are presumably valid and naively should be added together. Yet 
both explain to some extent the whole signal, in one  case for inclusive 
photons and in the other in the sub-domain of photon plus jet. It is not 
clear how to reconcile these, and we await further developments

\section*{References}
Most of the work described here can be found at
${\sf http://www-zeus.desy.de/physics/phch/conf/lp05\_eps05/}$ and
${\sf http://www-h1.desy.de/h1/www/general/home/intra\_home.html}$.
The prompt photon results are found at 
H1 collab., Eur.Phys.J {\bf{C38}} (2005) 437-445,
ZEUS collab., Phys Lett {\bf{B595}} (2004) 86-100 and
A D Martin et al., Eur.Phys.J {\bf C39} (2005) 155-161.

\end{document}